\documentclass{nature}
\usepackage{times}  
\usepackage{helvet}  
\usepackage{courier}  
\usepackage{hyperref}       
\usepackage{url}            
\usepackage{booktabs}       
\usepackage{amsfonts}       
\usepackage{nicefrac}       
\usepackage{microtype}      
\usepackage{subfigure}
\usepackage{multirow}
\usepackage{enumitem}
\usepackage{amsmath}
\usepackage{enumerate}
\usepackage{xcolor}
\usepackage{bm}
\usepackage{graphicx}
\usepackage{booktabs}
\usepackage{subfigure}
\usepackage{hyperref}
\usepackage{color}
\usepackage{colortbl}
\usepackage{chngpage}
\usepackage{import}
\usepackage{amssymb,amsfonts}
\usepackage{algorithmic}
\usepackage{textcomp}
\usepackage{flushend}
\usepackage{hhline}
\usepackage{multirow}
\usepackage{gensymb}
\usepackage{rotating}
\usepackage{caption}
\usepackage{xr}

\usepackage{tabularx}
\usepackage{hyperref}

\usepackage{threeparttable}
\usepackage{array}
\usepackage{lineno}

\usepackage[capitalise]{cleveref}

\graphicspath{{./figs/}}

\title{A Joint Representation Using Continuous and Discrete Features for Cardiovascular Diseases Risk Prediction on Chest CT Scans}

\author{Minfeng Xu$^{1,3*\dagger}$, 
Chen-Chen Fan$^{2*}$, 
Yan-Jie Zhou$^{1,3*}$,
Wenchao Guo$^{1,3}$,
Pan Liu$^{4,5}$,
Jing Qi$^{4,5}$,
Le Lu$^1$,
Hanqing Chao$^{1,3}$, 
Kunlun He$^{4,5\dagger}$}

\begin{document}
\maketitle              

\begin{affiliations}
 \item DAMO Academy, Alibaba Group
 \item Department of Electronic Engineering, Tsinghua University, Beijing, China
 \item Hupan Laboratory, Hangzhou, China
 \item Beijing Key Laboratory of Precision Medicine for Chronic Heart Failure, Beijing, China
 \item National Engineering Research Center for Medical Big Data Application Technology, Beijing, China

$*$ Asterisks indicate first coauthors, they contributed equally  \\
$\dagger$ Daggers indicate co-correspongding authors 
\end{affiliations}

\section*{Abstract}
\vspace{6pt}

\begin{abstract}
Cardiovascular diseases (CVD) remain a leading health concern and contribute significantly to global mortality rates. While clinical advancements have led to a decline in CVD mortality, accurately identifying individuals who could benefit from preventive interventions remains an unsolved challenge in preventive cardiology. Current CVD risk prediction models, recommended by guidelines, are based on limited traditional risk factors or use CT imaging to acquire quantitative biomarkers, and still have limitations in predictive accuracy and applicability. On the other hand, end-to-end trained CVD risk prediction methods leveraging deep learning on CT images often fail to provide transparent and explainable decision grounds for assisting physicians. In this work, we proposed a novel joint representation that integrates both discrete quantitative biomarkers and continuous deep features extracted from chest CT scans. Our approach initiated with a deep CVD risk classification model by capturing comprehensive continuous deep learning features while jointly obtaining currently clinical-established quantitative biomarkers (as discrete features) via segmentation models. In the feature joint representation stage, we use an instance-wise feature-gated mechanism to align the continuous and discrete features, followed by a soft instance-wise feature interaction mechanism fostering independent and effective feature interaction for the final CVD risk prediction. Our method substantially improves CVD risk predictive performance and offers individual contribution analysis of each biomarker, which is important in assisting physicians' decision-making processes. We validated our method on a public chest low-dose CT dataset (LDCT-NLST) and a private external chest standard-dose CT patient cohort of 17,207 CT volumes from 6,393 unique subjects, and demonstrated superior predictive performance, achieving AUCs of 0.875 and 0.843 for two cohorts, respectively.
\end{abstract}

\section*{Introduction}

Cardiovascular diseases (CVD) continue to pose a significant public health concern, accounting for 45\% of deaths in Europe~\cite{wilkins2017european} and 31\% in the United States~\cite{mozaffarian2016heart}. Despite a notable 7\% reduction in CVD mortality rates since 2005, largely attributed to advances in clinical screening and prevention strategies, heart disease remains one of the leading causes of death in the United States~\cite{d20191997}. The ongoing challenge is to identify asymptomatic patients who would benefit most from preventive measures, such as initiating statin therapy~\cite{ridker2008rosuvastatin} and others, in a convenient, accessible, accurate, and non-invasive manner. This is a crucial aspect of effective preventive cardiology patient management.

Several guidelines endorse different CVD risk prediction models. The 2010 guidelines~\cite{greenland20102010} from the American College of Cardiology / American Heart Association (ACC / AHA) suggest the use of the Framingham Risk Score (FRS)~\cite{wilson1998prediction,eichler2007prediction,d2008general}, a well-established and thoroughly validated model that considers traditional risk factors such as cholesterol levels and blood pressure, etc. In contrast, the 2016 European guidelines favor the Systematic Coronary Risk Evaluation (SCORE) model~\cite{piepoli2016guidelines}. The 2016 Chinese guidelines introduce the China-PAR model, specifically designed to estimate the 10-year risk for atherosclerotic cardiovascular disease~\cite{yang2016predicting,liu2018predicting}. However, these recommended risk prediction models rely on a limited set of predictors, which may result in suboptimal risk management performances on various patient populations.

Recent studies have begun to utilize CT scans to extract quantitative imaging biomarkers, demonstrating good potential for surpassing the predictive performance of multivariate FRS~\cite{pickhardt2020automated}. With the developments of deep learning in medical imaging, previous work explored integrating CT scans with deep learning algorithms to automatically obtain CT quantitative biomarkers related to CVD risk~\cite{xu2023ai,eng2021automated,zeleznik2021deep}. The methods specialized in multivariate analysis are designed for CVD risk prediction and can contain considerable interpretability. However, they may often overlook the interactions among different factors, and due to restrictive modeling assumptions and a limited number of predictors, these methods tend to attain only moderate predictive performance~\cite{siontis2012comparisons}. On the other hand, some studies~\cite{van2019direct,chao2021deep} have directly applied deep learning to predict CVD risk from chest CT scans. While these methods can greatly improve the predictive capacity and accuracy, they provide little interpretable insight into the model's decision-making process.

In this study, we propose a simple yet effective information fusion method, called DeepCVD, that extracts continuous deep features and calculates discrete quantitative biomarkers from CT scans. These features are subsequently integrated using a joint representation module that fosters interaction, ultimately enhancing the precision of CVD risk prediction and enabling the analysis of each feature's contribution to the model's decision-making process. In the feature extraction phase, discrete features are generated from image segmentation models as CT quantitative biomarkers, crafted based on current clinical knowledge to improve the model’s predictive accuracy and applicability. Meanwhile, continuous deep image features derived from a trained deep CVD classifier encompass rich higher-order semantic contextual information pertinent to CVD, albeit challenging to quantify. We propose a joint representation module to fully utilize the complementary information of these two channels of features for superior CVD risk prediction. Specifically, we employ an instance-wise feature gating mechanism (IFGM) to align the densities and dimensions of the input continuous and discrete features. We apply a Gated Residual Network (GRN)~\cite{lim2021temporal}, a flexible and non-linear operation, to obtain more robust feature embeddings. Inspired by the multi-head attention mechanism~\cite{vaswani2017attention}, we have designed a soft instance-wise feature interaction mechanism (SIFIM), facilitating thorough interactions among features across diverse subspaces and capturing their intricate interrelations for joint representation learning. Importantly, this strategy preserves the autonomy of each input feature, enabling physicians to understand their respective contributions to the model’s decisions easily. Furthermore, our feature interaction technique allows the model to adaptively output the contribution of each feature for various CVD, aligning closely with practical clinical settings. Our approach has achieved competitive performance using both public datasets and a private external validation patient cohort.

\begin{figure*}[htbp]
\renewcommand{\figurename}{Fig.}
\centering
\includegraphics[width=\textwidth]{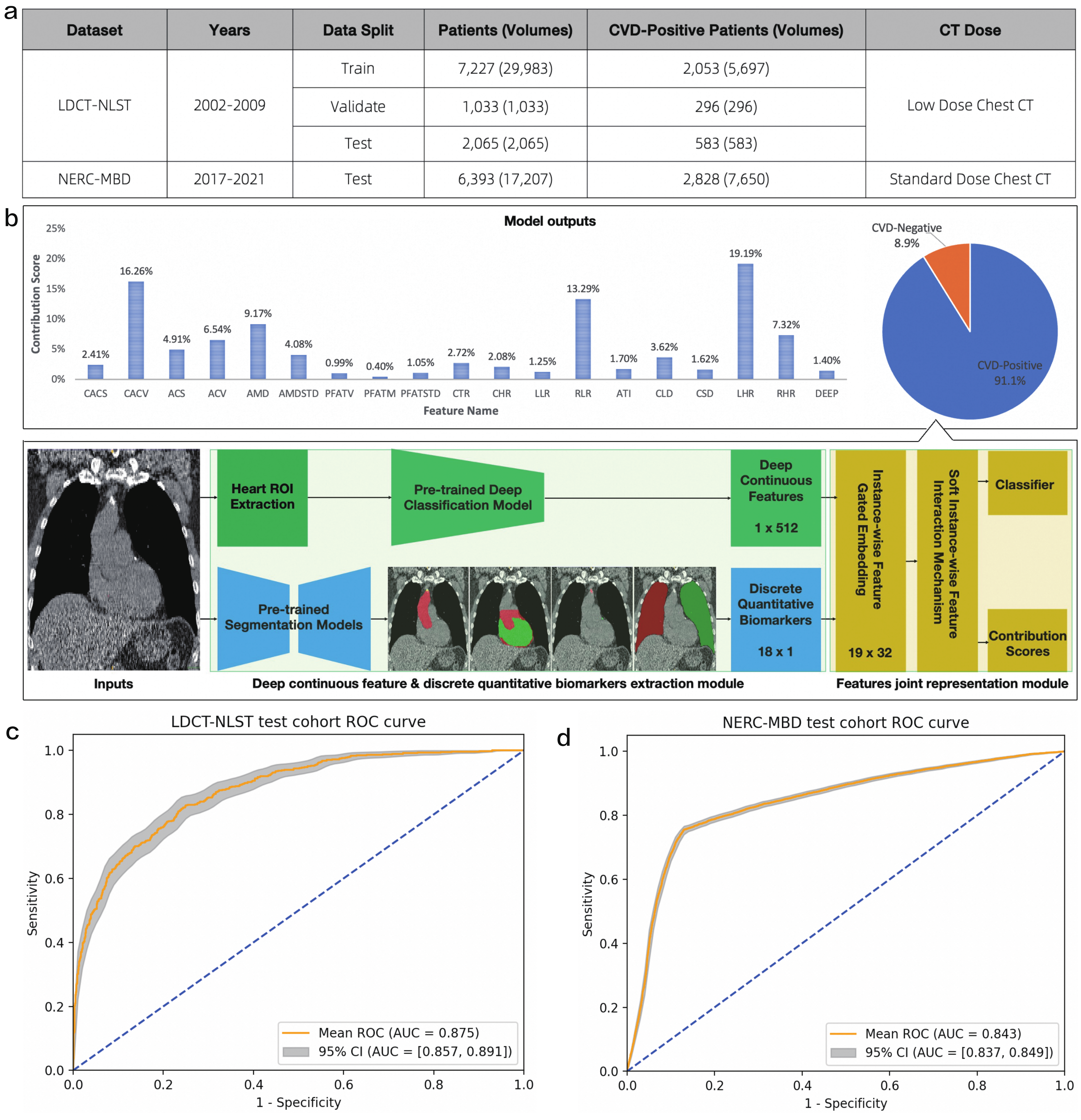}
\caption{\textbf{\textbar~Overview of the proposed DeepCVD framework, the training and testing cohorts.} \textbf{a.} Pubilic LDCT-NLST and the external standard-dose chest CT (NERC-MBD) cohorts. \textbf{b.} Schematic overview of DeepCVD. It takes chest CT as input and outputs the probability of CVD risk, and the individual contribution score of each biomarker. The DeepCVD framework consists of two stages: the first stage involves the extraction of deep continuous features and discrete CT quantitative biomarkers, while the second stage entails features joint representation followed by CVD risk prediction. \textbf{c.} ROC curves of CVD risk prediction on LDCT-NLST and NERC-MBD testing cohorts.}
\label{fig:overview}
\end{figure*}

\section*{Results}

\subsection{Dataset overview.} In this retrospective study, a total of two cohorts were used to develop and evaluate the performance of DeepCVD (\cref{fig:overview}a). Specifically, the National Lung Screening Trial (LDCT-NLST) cohort comprised 10,395 patients with 33,413 volumes, whereas the National Engineering Research Center for Medical Big Data (NERC-MBD) cohort included 6,393 patients with 17,207 volumes.

The LDCT-NLST cohort is a significant dataset intended to evaluate the effectiveness of low-dose chest CT scans in lung cancer screening. In this trial, it also collects information related to the CVD of subjects, making the dataset applicable for research in CVD risk prediction~\cite{chao2021deep, van2019direct}. Each subject undergoes one to three CT examinations, each producing multiple CT scans employing different CT reconstruction kernels. It has 33,413 CT volumes from 10,395 subjects, and each subject is labeled as either CVD-Positive or CVD-Negative. A subject is considered CVD positive if any cardiovascular abnormalities are reported in their CT screening examinations, or if the subject dies of CVD. CVD-negative subjects have no history of CVD during the clinical trial, no cardiovascular abnormalities reported in any CT scans, and do not die from circulatory system diseases.

The NERC-MBD cohort comprises 17,207 standard-dose chest CT volumes from 6,393 subjects. Of these, 7,650 CT volumes are from 2,828 CVD-positive subjects, and 9,557 CT volumes are from 3,565 CVD-negative subjects. A subject labeled as CVD-positive is identified based on the diagnosis report from the Department of Cardiology. Subjects diagnosed with acute cardiovascular disease have CT scans from the year prior to diagnosis collected, while those diagnosed with chronic cardiovascular disease have CT scans from the two years prior to diagnosis collected. Subjects without any cardiovascular abnormalities in their CT examination reports or clinical conclusions are categorized as CVD-negative. It is important to note that this cohort includes subjects with 16 types of cardiovascular diseases, and detailed information is summarized in Extended Data \cref{fig:cvd_diseases_subtypes}.

\subsection{Development of joint representation CVD risk prediction system.} We developed the continuous and discrete features joint representation CVD risk prediction system (DeepCVD) in the LDCT-NLST cohort (\cref{fig:overview}a). We adopt the three random subsets generated by the prior study~\cite{chao2021deep} entirely, with the training set accounting for 70\% (7,268 subjects), the validation set 10\% (1,042 subjects), and the independent test set 20\% (2,085 subjects). We found that the LDCT-NLST cohort contained some data that did not meet the requirements of our model, such as incomplete field of view (FOV), missing slices, and severe cardiac artifacts. Two radiologists cleansed the dataset and removed the problematic subjects, resulting in 10,325 subjects. The training set comprises 7,227 subjects, the validation set comprises 1,033 subjects, and the test set includes 2,065 subjects. Our goal was to achieve accurate CVD risk prediction using chest CT scans while providing physicians with reliable bases for model decisions.

We designed a two-stage system. In the first stage, eighteen discrete CT quantitative biomarkers based on physician’s insights, along with continuous deep features, were extracted from the LDCT-NLST training set and validation set. The second stage focused on training the joint representation CVD risk prediction model based on the discrete quantitative biomarkers and continuous deep features. The process began with feature alignment through an instance-wise feature-gated mechanism to obtain independent embedding vectors, followed by a soft instance-wise feature interaction mechanism to conduct feature interactions at the instance level and to compute attention weights for each instance feature to achieve joint representation. Finally, the model outputted prediction outcomes and contribution scores for each instance feature (\cref{fig:overview}b). The CVD risk prediction results of our model were evaluated on the LDCT-NLST testing cohort and the NERC-MBD cohort (\cref{fig:overview}c).

\begin{table}[htbp]
\centering
\caption{\textbf{Quantitative evaluation of DeepCVD with different methods on LDCT-NLST testing cohort.}}
\begin{threeparttable}
\label{tab:table_nlst}
{\scriptsize
\begin{tabularx}{\textwidth}{ccccccc} 
\toprule
Methods        & Accuracy (95\% CI)  & Sensitivity (95\% CI) & Specificity (95\% CI) & F1 Score (95\% CI)  & AUC (95\% CI)       & p-value   \\ 
\hline
DeepCVD$^m$           & 0.833 (0.816-0.848) & 0.616 (0.576-0.655)   & 0.918 (0.903-0.931)   & 0.675 (0.642-0.708) & 0.875 (0.857-0.891) & -         \\
Tri2D-Net$^{*c}$      & 0.819 (0.802-0.837) & 0.485 (0.444-0.525)   & 0.952 (0.940-0.962)   & 0.603 (0.565-0.640) & 0.869 (0.850-0.886) & ~~~0.118  \\
ResNet34$^c$       & 0.796 (0.779-0.814) & 0.543 (0.502-0.583)   & 0.896 (0.880-0.911)   & 0.601 (0.563-0.634) & 0.844 (0.825-0.861) & $< 0.001$   \\
nnUNet-J$^c$   & 0.824 (0.806-0.840) & 0.599 (0.559-0.638)   & 0.913 (0.898-0.926)   & 0.658 (0.621-0.691) & 0.874 (0.856-0.890) & ~~~0.034  \\
ViT-B$^c$    & 0.651 (0.629-0.669) & 0.560 (0.519-0.600)   & 0.686 (0.662-0.709)   & 0.475 (0.442-0.507) & 0.676 (0.650-0.701) & $< 0.001$   \\
nnFormer-J$^c$ & 0.788 (0.769-0.805) & 0.692 (0.653-0.728)   & 0.825 (0.805-0.844)   & 0.648 (0.616-0.678) & 0.837 (0.817-0.856) & $< 0.001$   \\
Xgboost$^d$        & 0.771 (0.755-0.788) & 0.264 (0.230-0.301)   & 0.971 (0.961-0.978)   & 0.394 (0.352-0.437) & 0.835 (0.817-0.853) & $< 0.001$   \\
\bottomrule
\end{tabularx}
}
\begin{tablenotes}
\scriptsize
\item[]
Note: ``*'' indicates that Tri2D-Net is directly used as an open-source model trained on the LDCT-NLST training cohort, all other models are trained from scratch based on the LDCT-NLST training cohort. ``J'' indicates that we have incorporated a classification head into the classic segmentation networks, thus achieving multitask joint training to enhance CVD risk prediction performance. Detailed information about comparison methods can be found in the Methods Section. ``m'' indicates a model that jointly represents discrete quantification features and continuous deep features. ``c'' represents an end-to-end deep learning model. ``d'' represents a model that uses only discrete quantification features.

\end{tablenotes}
\end{threeparttable}
\end{table}

\subsection{Performance on the internal LDCT-NLST testing cohort.} The proposed DeepCVD was used to identify subjects with CVDs in the  LDCT-NLST testing cohort, which consisted of 2,065 patients (583 patients CVD-Positive and 1,482 patients CVD-Negative). The quantitative performance was summarized in \cref{tab:table_nlst}, and the receiver operating characteristic curves (ROCs) of multiple methods were shown in \cref{fig:overview}c. DeepCVD achieved an area under the curve (AUC) of 0.875 (95\% Confidence Interval (CI), 0.857-0.891), a sensitivity of 0.616 (95\% CI, 0.576-0.655), and a specificity of 0.918 (95\% CI, 0.903-0.931). From \cref{tab:table_nlst}, it was evident that models using our designed quantitative biomarkers or directly using prior knowledge, such as pericoronary calcification and epicardial fat, whether Xgboost~\cite{chen2016Xgboost}, Tri2D-Net~\cite{chao2021deep}, or our proposed DeepCVD, demonstrated significantly better specificity compared to models like ResNet34~\cite{he2016deep} that did not use prior knowledge. However, the complex encoding of prior knowledge can impair the ability to suppress false positives (non-encoded Xgboost specificity is 0.971, while encoded Tri2D-Net and DeepCVD had specificities of 0.952 and 0.918, respectively). By combining quantitative biomarkers and continuous deep features, DeepCVD increased the sensitivity by 27.0\% and the F1 score by 11.9\% compared to Tri2D-Net, at the expense of slightly reduced specificity.

\begin{figure*}[htbp]
\renewcommand{\figurename}{Fig.}
\centering
\includegraphics[width=\textwidth]{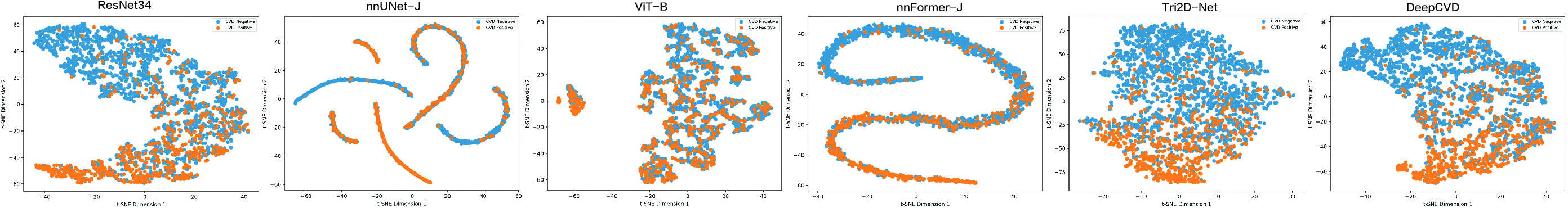}
\caption{\textbf{\textbar~Embedding visualization of different methods for LDCT-NLST testing cohort.} The embeddings displayed from left to right are: ResNet34, nnUNet-J, ViT-B, nnFormer-J, Tri2D-Net, and DeepCVD.}
\label{fig:t_sne}
\end{figure*}

To further present the quality of the embeddings of different methods in the LDCT-NLST testing cohort, t-SNE~\cite{van2008visualizing} was used to visualize the embeddings of different methods. In \cref{fig:t_sne}, the color of the nodes corresponded to CVD-Positive and CVD-Negative, verifying the discriminative power of the method. As indicated in \cref{fig:t_sne}, DeepCVD and Tri2D-Net can achieve more compact and separated clusters compared with other methods that do not incorporate prior insights.

\subsection{Performance on the external standard-dose chest CT testing cohort.} To assess the generalizability of our proposed DeepCVD, we directly applied the model trained on the LDCT-NLST dataset to the NERC-MBD standard-dose chest CT dataset. DeepCVD achieved an AUC of 0.843 (95\% CI, 0.837-0.849), accuracy of 0.819 (95\% CI, 0.813-0.824), sensitivity of 0.756 (95\% CI, 0.746-0.765), and specificity of 0.870 (95\% CI, 0.863-0.876). These results showed significant performance improvement (p $< 0.001$) compared to the previous state-of-the-art approach Tri2D-Net~\cite{chao2021deep} (accuracy increased by 9.8\%, sensitivity by 28.4\%, F1 score by 17.1\%, and AUC by 5.9\%, while the specificity remained almost the same). In comparison, other methods had inferior performance that was statistically significant (all p $< 0.001$). Detailed evaluation performance was summarized in \cref{tab:table_mbdc}.

Although the specificity of DeepCVD was not the highest, all other quantitative performance metrics were the best. Additionally, compared to other methods, DeepCVD demonstrated the most stable performance on both the internal LDCT and external standard-dose chest CT dataset (AUC: internal 0.875, external 0.843; ACC: internal 0.835, external 0.819), showcasing DeepCVD’s generality and accuracy holding well in this large-scale external testing cohort.

\begin{table}[htbp]
\centering
\caption{\textbf{Quantitative evaluation of DeepCVD with different methods on NERC-MBD testing cohort.}}
\label{tab:table_mbdc}
{\scriptsize
\begin{tabularx}{\textwidth}{ccccccc} 
\toprule
Methods   & Accuracy (95\% CI)  & Sensitivity (95\% CI) & Specificity (95\% CI) & F1 Score (95\% CI)  & AUC (95\% CI)       & p-value   \\ 
\hline
DeepCVD$^m$      & 0.819 (0.813-0.824) & 0.756 (0.746-0.765)   & 0.870 (0.863-0.876)   & 0.788 (0.780-0.795) & 0.843 (0.837-0.849) & -         \\
Tri2D-Net$^{*c}$ & 0.746 (0.739-0.752) & 0.589 (0.578-0.600)   & 0.872 (0.865-0.878)   & 0.673 (0.664-0.682) & 0.796 (0.790-0.803) &  $< 0.001$  \\
ResNet34$^c$  & 0.796 (0.791-0.803) & 0.705 (0.695-0.715)   & 0.869 (0.863-0.876)   & 0.755 (0.747-0.763) & 0.832 (0.826-0.839) &  $< 0.001$  \\
nnUNet-J$^c$    & 0.708 (0.701-0.714) & 0.747 (0.737-0.756)   & 0.676 (0.667-0.686)   & 0.694 (0.686-0.702) & 0.780 (0.773-0.787) & $< 0.001$  \\
ViT-B$^c$     & 0.643 (0.635-0.649) & 0.720 (0.710-0.730)   & 0.580 (0.570-0.590)   & 0.642 (0.633-0.650) & 0.716 (0.708-0.724) &  $< 0.001$  \\
nnFormer-J$^c$  & 0.726 (0.720-0.733) & 0.751 (0.741-0.761)   & 0.706 (0.697-0.715)   & 0.709 (0.701-0.717) & 0.795 (0.788-0.802) &  $< 0.001$  \\
Xgboost$^d$   & 0.679 (0.672-0.686) & 0.369 (0.358-0.380)   & 0.928 (0.922-0.933)   & 0.505 (0.494-0.517) & 0.820 (0.813-0.826) &  $< 0.001$  \\
\bottomrule
\end{tabularx}
}
\end{table}

\subsection{Feature contribution analysis and visualization.} 
\begin{figure*}[htbp]
\renewcommand{\figurename}{Fig.}
\centering
\includegraphics[width=\textwidth]{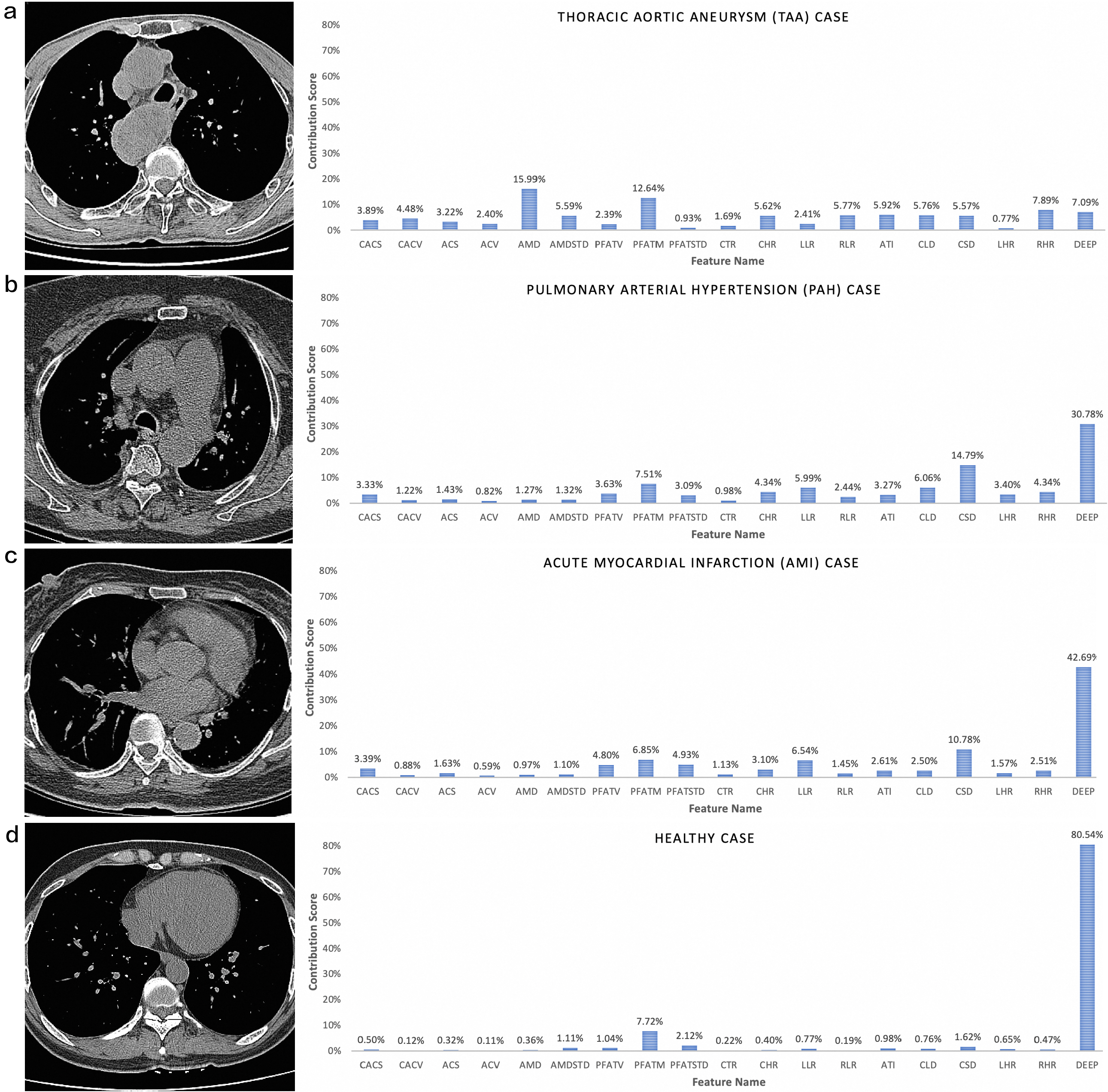}
\caption{\textbf{\textbar~Visual representation of the contribution score for different features calculated by DeepCVD, depicted across four distinct scenarios.} \textbf{a.} An example of a thoracic aortic aneurysm visible on CT volume, where quantitative biomarkers describing the shape of the aorta (AMD, AMDSTD, and ATI) demonstrate significant responsiveness in the decision-making process. \textbf{b.} An instance of pulmonary hypertension is visible on CT volume. Although DeepCVD does not incorporate quantitative biomarkers associated with the pulmonary arteries, it exhibits a strong response to quantitative biomarkers related to the heart in the assessment. \textbf{c.} An example of acute myocardial infarction not visibly apparent on CT volume, where quantitative biomarkers characterizing the heart show responsiveness, and the contribution of deep features to the decision-making process is increasing. \textbf{d.} Depicts a CVD-Negative case, in which the quantitative biomarkers we designed exhibit minimal responsiveness, and deep features predominantly drive the decision-making process.}
\label{fig:contribution_scores}
\end{figure*}

To interpret the contribution of each instance-wise feature in the decision-making process of CVD risk prediction, we have categorized four scenarios to demonstrate the contribution scores of the features learned by DeepCVD (\cref{fig:contribution_scores}). Detailed extraction of quantitative CT biomarkers is illustrated in the Methods Section.

(i) Evidence that is directly visible in the CT scans and can diagnose certain CVD, is also included in the quantitative biomarkers. The corresponding biomarkers usually make the main decision of our model. \cref{fig:contribution_scores}a shows a CT volume with the thoracic aortic aneurysm (TAA). Within the learned contribution scores, we find that the biomarkers describing the shape of the aorta (AMD, AMDSTD, and ATI) accounted for 27.5\%, playing a decisive role, while deep features contributed only 7.1\%.

(ii) Evidence that is directly visible in the CT volume, which can diagnose certain CVD, but is not included in the quantitative biomarkers, usually results in the main decision of our model being made jointly by deep features and relevant biomarkers. \cref{fig:contribution_scores}b shows an example of a case of pulmonary arterial hypertension (PAH), where the diameter of the pulmonary artery is noticeable enlarged and exceeds that of the ascending aorta, corresponding to the typical imaging symptoms of PAH. The information about the pulmonary artery diameter is not included directly in our quantitative biomarkers, but PAH will also bring indirect signs of pulmonary texture and heart morphology. The biomarkers related to heart morphology (CHR, CLD, and CSD) accounted for 25.2\%, the biomarkers related to pulmonary textures (LLR, RLR, LHR, and RHR) accounted for 16.2\%, and the deep features contributed 30.8\%, forming good complementary for each other.

(iii) Evidence that indirectly suggested the presence of certain CVD visible in CT scans. Deep features mainly make the decision of our model, but other related biomarkers also had corresponding decision contributions. \cref{fig:contribution_scores}c shows a CT volume from a patient with acute myocardial infarction (AMI) six months ago, from which we can observe that the decision proportion of deep features accounted for 42.7\%. The proportion of biomarkers related to heart morphology (CHR, CLD, and CSD) accounted for 16.4\%, biomarkers related to pericardial fat (PFATV, PFATM, and PFATSTD) accounted for 16.6\%, and biomarkers related to coronary calcification (CACS and CACV) accounted for 4.3\%.

(iv) CT scans did not have imaging evidence to diagnose a certain CVD. Deep features almost entirely decided on our model. \cref{fig:contribution_scores}d shows a CT volume from a healthy individual, where we can see that deep feature contributed for 80.5\%. 

To further explore whether the distribution of the contribution score to learned characteristics exhibits characteristics associated with different diseases, we presented additional illustrations in various CVDs (see Extended Data \cref{fig:contribution_score_s_1} and Extended Data \cref{fig:contribution_score_s_2}).

\section*{Discussion}
It is essential to identify asymptomatic patients at risk of CVD who could benefit greatly from preventive measures. While significant progress has been made in this area, challenges such as suboptimal prediction performance or a lack of interpretable evidence from model outputs make it difficult to use these models in clinical settings. Our DeepCVD can simultaneously represent continuous deep features and discrete quantitative biomarkers to address these issues. This framework fully utilizes the complementary strengths of prior knowledge-based discrete quantitative biomarkers and rich, CVD-relevant continuous deep features. The joint representation of features enables comprehensive interaction and integration between features, significantly enhancing CVD risk prediction performance. Furthermore, our joint representation is carried out at an instance-wise feature level, providing adaptive contribution scores for each biomarker and deep feature in the model decision-making process, offering physicians reliable predictive results.

This study thoroughly validates the effectiveness of DeepCVD. It successfully combines the advantages of discrete and continuous features, improves CVD risk prediction performance, and demonstrates strong generalizability. Multisource feature fusion techniques are also widely used in the medical domain, as feature concatenation is a commonly adopted and effective method for joint feature representation~\cite{kiela2014learning}. In the first stage, we concatenate continuous and discrete features in this manner and then feed them into the second stage features joint representation module. We observe an increase in AUC from 0.844 to 0.862 compared to directly using a CVD risk classifier with ResNet34. However, this mode of feature interaction does not fully leverage the respective advantages of quantitative discrete biomarkers and deep continuous features, with specificity increasing from 0.896 to 0.956, while sensitivity decreased from 0.543 to 0.440 (as indicated in \cref{tab:table_nlst} and \cref{fig:ablation_study}c). Furthermore, although our model is trained on LDCT-NLST, it demonstrates remarkable generalization to standard-dose chest CT scans as shown in \cref{tab:table_mbdc}. Moreover, our joint representation module outperformed models like Xgboost, even when only discrete features were used, demonstrating the benefits of effective feature interaction for enhancing model performance (shown in \cref{fig:ablation_study}d).

We have also discovered that incorporating prior knowledge, either through direct predictions based on CT quantitative biomarkers using Xgboost or by integrating them into deep models such as Tri2D-Net, effectively improves the model's specificity. Tri2D-Net incorporated pericardial fat and calcification as strong constraints in the prediction model. In the real world, not all CVDs are strongly associated with just two biomarkers, which reduces the model’s sensitivity and its ability to generalize to real-world scenarios. As shown in \cref{fig:t_sne}, although Tri2D-Net also demonstrates high-quality embeddings across the LDCT-NLST test set, its performance significantly drops on the external test set (see \cref{tab:table_mbdc}). This is mainly because the external test set includes many subjects with cerebral infarction (refer to Extended Data \cref{fig:cvd_diseases_subtypes}). Consequently, using CAM~\cite{selvaraju2017grad} for model interpretability also becomes less meaningful. In contrast, our model aligns and flexibly processes discrete and continuous features through the instance-wise feature-gated mechanism to generate more robust embeddings. Then, the soft instance-wise feature interaction mechanism achieves thorough interaction and fusion of features while maintaining relative independence. This ensures that our model optimally balances specificity and sensitivity (as seen in \cref{tab:table_nlst} and \cref{tab:table_mbdc}), and automatically learns the relationships between different CVDs and various biomarkers and deep features (as seen in \cref{fig:contribution_scores}). This informs physicians about the role of each feature in the decision-making process and opens up possibilities to discover new clinical biomarkers in the CVD domain.

A recent study utilized a mature and fully automated abdominal CT-based algorithm with predefined metrics to quantify aortic calcification, muscle density, the ratio of visceral to subcutaneous fat, liver fat, and bone mineral density for assessing CVD risk~\cite{pickhardt2020automated}. The research demonstrated that the multivariate combination of CT biomarkers could effectively enhance CVD risk prediction performance over traditional risk factors. For example, the combination of four CT-based quantitative biomarkers—aortic calcification, muscle density, the ratio of visceral to subcutaneous fat, and liver fat—resulted in a 2-year AUC of 0.817 (95\% CI 0.768-0.866). Although this study used a different retrospective cohort and abdominal non-contrast CT, the results of the two studies both demonstrated the potential CVD risk prediction value of harnessing the rich biometric tissue data embedded within all body CT scans that typically go unused in routine practice. However, this approach requires predefined CT biomarkers, and the combination of these biomarkers to enhance CVD risk prediction performance needs to be validated through repeated experiments. Our approach goes a step further by using deep features to represent those CT biomarkers that cannot or have not yet been predefined, thereby enhancing CVD risk prediction performance. Furthermore, through a unique design, our model can output the contribution of each feature in the decision-making process. This not only provides more insights for doctors but also suppresses those CT biomarkers that do not aid in CVD risk prediction.

While DeepCVD achieved significant results on both the LDCT-NLST and NERC-MBD testing cohorts with interpretable predictions, this study does have certain limitations. CVD encompasses a range of conditions affecting various organs and tissues, including the brain and heart. However, the LDCT-NLST and NERC-MBD testing cohorts consist of a limited range of diseases with an uneven distribution of chest CT quantities for each disease. Additionally, our model trained on the chest LDCT, focuses solely on discrete CT quantitative biomarkers and deep features within the chest region. Some CVDs may not present direct or indirect signs on chest CT, or they may not have affected the organs or tissues in the chest yet, leading to lower prediction performance. In the NERC-MBD testing cohort, we observed that the prediction performance for diseases such as cerebral infarction and occlusion of the precerebral artery is lower than for diseases related to the heart and major blood vessels, such as ischemic heart disease (Extended Data \cref{fig:cvd_diseases_subtypes}), indicating that there is room for improvement in our method. Lastly, while we gained insight into the role of each feature in the decision-making process through the learned contribution score (shown in \cref{fig:contribution_scores}), we have not yet established a direct link between biomarkers and specific CVDs through the contribution score, especially when deep features predominantly drive the model’s decisions.

In clinical practice, the fusion of imaging information with clinical information can result in increased accuracy, mode informative clinical decision-making, and improved patient outcomes~\cite{huang2020fusion}. Our DeepCVD can seamlessly integrate into clinical information. In future work, we aim to expand the discrete features from CT quantitative biomarkers to multimodal biomarkers, incorporating laboratory indicators closely linked to CVDs such as blood pressure, cholesterol, and smoking history into the model to enhance predictive performance. Additionally, we plan to establish connections between biomarkers and specific CVDs through contribution scores, providing actionable guidance for physicians in subsequent diagnoses and achieving greater clinical value.

\section*{Methods}
This study aims to accurately predict CVD risk using chest CT scans and provide physicians with reliable information for model decisions. Our method is based on two findings and assumptions. First, while well-known quantitative biomarkers from CT scans, such as the coronary artery calcium score, are considered risk factors for CVD~\cite{pickhardt2020automated,iacobellis2022epicardial}, we believe that many undefined or difficult-to-quantify biomarkers related to CVD have not been fully exploited within chest CT scans.  Fully utilizing these biomarkers can effectively improve the performance of CVD risk prediction. Second, the relevance and contributions of biomarkers should vary across different CVDs, requiring these features to interact sufficiently to address the diversity of CVDs while also maintaining relative independence so that the model can adaptively output the contribution of each biomarker in decision-making.

We present a new pipeline for CVD risk prediction, as depicted in \cref{fig:overview}b. The pipeline consists of two main stages. In the first stage, we extract discrete CT quantitative biomarkers and continuous deep features. We generate $N$ discrete quantitative biomarkers (where $N=18$ in this study) based on clinical insights and four pre-trained segmentation models. We use deep features obtained from a pre-trained deep CVD risk classifier to comprehensively represent these undiscovered or difficult-to-quantify biomarkers, which serve as continuous deep features. The second stage primarily involves the joint representational learning of discrete quantitative biomarkers and continuous deep features. It begins with feature alignment through an instance-wise feature-gated mechanism to obtain independent embedding vectors, followed by a soft instance-wise feature interaction mechanism to conduct feature interactions at the instance level and compute attention weights for each instance feature to achieve joint representation. Finally, the model outputs prediction outcomes and contribution scores for each instance feature.

\subsection{Deep continuous feature and discrete biomarker extraction}

\begin{figure*}[t]
\renewcommand{\figurename}{Fig.}
\centering
\includegraphics[width=\textwidth]{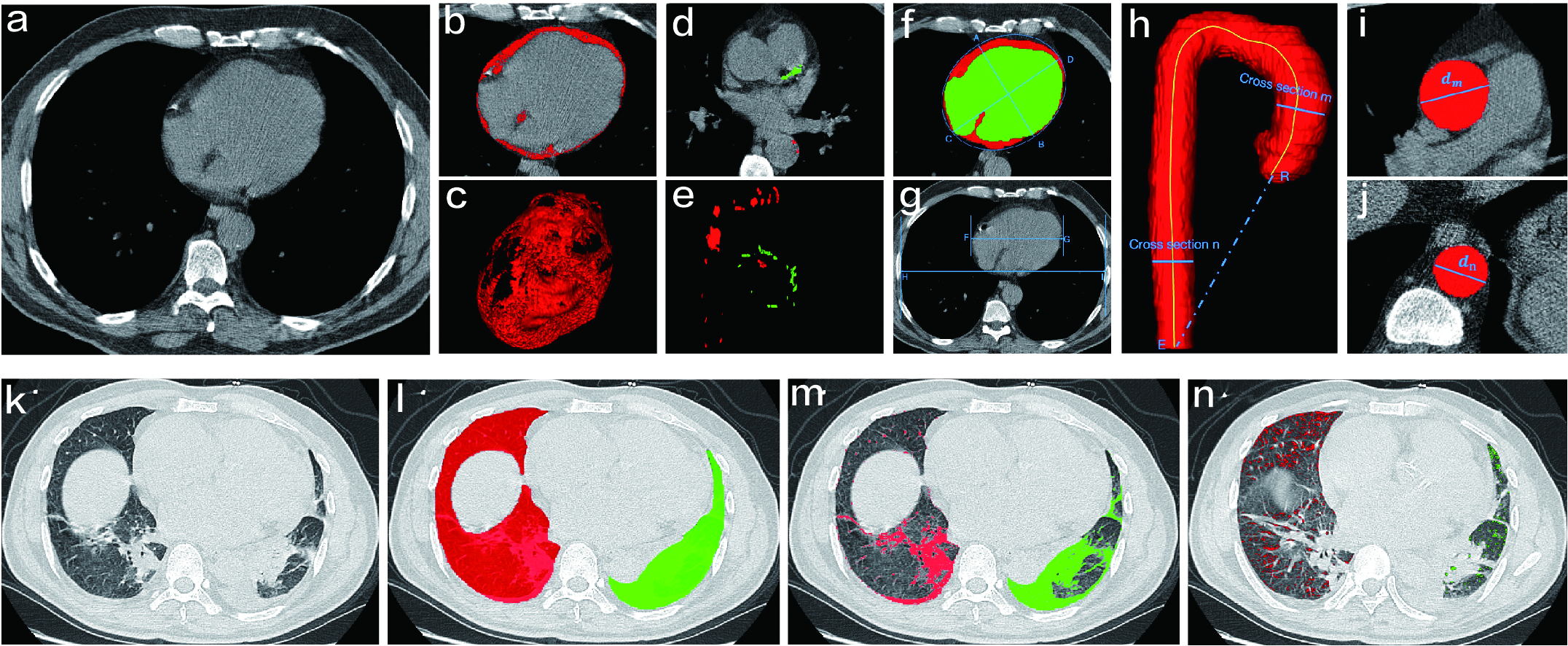}
\caption{\textbf{\textbar~Discrete CT quantitative biomarkers visualization.} \textbf{a - j.} visually illustrate discrete CT quantitative biomarkers related to the heart and vasculature. \textbf{k - n.} represent discrete CT quantitative biomarkers related to the lungs. \textbf{a.} The input CT scan under the mediastinal window. \textbf{b.} Segmentation result of pericardial fat in the axial view. \textbf{c.} 3D visualization of pericardial fat. \textbf{d.} Segmentation results for coronary artery calcification (green) and aortic calcification (red) in the axial view. \textbf{e.} 3D visualization of coronary artery and aortic calcification. \textbf{f.} Axial view showing the segmentation results for cardiac chambers (green) and pericardium (red), as well as a schematic representation of the calculation of cardiac long and short axes. \textbf{g.} A schematic representation of calculating the cardiothoracic ratio. \textbf{h.} 3D visualization and the centerline of the thoracic aorta. \textbf{i.} Cross-sectional view illustrating the segmentation at the ascending aorta. \textbf{j.} Cross-sectional view depicting the segmentation at the descending thoracic aorta. \textbf{k.} The input CT scan under the lung window. \textbf{l.} Lung segmentation results are displayed in the axial view. \textbf{m.} Segmentation outcome of high attenuation regions within the lungs. \textbf{n.} Segmentation results of low attenuation regions within the lungs.}
\label{fig:discrete_feature_extraction}
\end{figure*}

\subsection{Continuous deep features extraction.} The extraction of deep continuous features relies on a pre-trained CVD risk prediction model, which has been trained on the LDCT-NLST training set. The model is based on a progressive coarse-to-fine framework. In the coarse stage (heart localization), a lightweight cardiac segmentation model is applied to localize the volume of the region of interest in the heart. The input chest LDCT scan is cropped into a smaller volume to reduce the influence of irrelevant content noise and conserve computational resources. Moving to the fine stage, the cropped volume is used to train a classifier based on ResNet34~\cite{he2016deep}, tasked with distinguishing between CVD-Positive and CVD-Negative cases. Ultimately, the outputs from the layer preceding the final fully connected layer of ResNet34 are extracted. These high-dimensional embeddings are rich in features relevant to CVDs, denoted as $x_1$ ($x_1 \in \mathbb{R}^{1 \times D}$, here $D=512$), and serve as continuous deep features of our proposed method.

\subsection{Discrete CT quantitative biomarkers extraction.} The CT quantitative biomarkers extraction process utilizes four specialized body part segmentation models, each trained on the 400 internal chest CT scans. These models are designed to segment different anatomical structures, including the heart chambers and pericardium, aortic and coronary calcium, aortic structure, and left and right lungs. The foundational architecture for these models is MedFormer~\cite{gao2022data}. These four segmentation models are applied to segment the whole LDCT-NLST dataset, with the results being reviewed by two radiologists and revised if necessary. The overall failure rate is less than 0.5$\%$. \cref{fig:discrete_feature_extraction} visually showcases the results obtained from these fully automated body part segmentation models.

We have established stable quantitative measures for each tissue composition based on the automated segmentation results without additional learning or adjustment. A total of $N$ quantitative biomarkers are calculated, with certain biomarkers such as the coronary artery calcium score (CACS) and the cardiothoracic ratio (CRT) having established associations with CVD and mortality in previous research~\cite{pickhardt2020automated,iacobellis2022epicardial,girardi2021aortic, hemingway1998cardiothoracic,dey2012epicardial}. Others are characterized by physicians using the results of the four segmentation models. All of these biomarkers are scalar and are symbolized as $x_i$ (where $i \in [2, N+1]$). 

(i) Based on the segmentation results of the pericardium, three quantitative biomarkers related to pericardial fat are calculated: PFATV, PFATM, and PFATSTD. PFATV primarily serves to quantify the volume of pericardial fat. Initially, we used the result of pericardial segmentation (here marked in red in~\ref{fig:discrete_feature_extraction}f) to identify the location of the pericardium on the CT scan. Subsequently, we further delineate the regions of pericardial fat (as depicted in~\ref{fig:discrete_feature_extraction}b) within the Hounsfield Unit (HU) range of [-190HU, -30HU]~\cite{dey2012epicardial}. Following this, we count the number of voxels within the 3D pericardial fat mask (illustrated in~\ref{fig:discrete_feature_extraction}c), and by multiplying this count with the voxel volume ($x$ resolution $\times$ $y$ resolution $\times$ $z$ resolution), we derive the value for PFATV, expressed in $mm^3$. Finally, we compute the mean and standard deviation of the attenuation intensity values for all voxels within the pericardial fat regions, denoted as PFATM and PFATSTD, respectively. 

(ii) Based on the segmentation results of calcifications (Coronary Artery Calcification and Thoracic Aorta Calcification), four calcification-related quantitative biomarkers are calculated: CACS, CACV, ACS, and ACV. CACS and ACS are primarily derived from an area perspective, calculated according to the Agatston Score~\cite{agatston1990quantification} for coronary and thoracic aortic calcification scores, respectively. CACV and ACV, on the other hand, are calculated from a volumetric perspective, quantifying the volume of coronary calcification and thoracic aortic calcification, respectively. First, the calcification segmentation results are refined, with only regions with CT attenuation intensity greater than or equal to 130HU retained as the final calcification segmentation mask. Then, calculating calcification scores, the results are uniformly reconstructed to 3mm before computing CACS and ACS to minimize the impact of different CT slice thicknesses. In the calculation of calcification volume, the number of voxels within the 3D coronary calcification mask (green areas in \cref{fig:discrete_feature_extraction}d and \cref{fig:discrete_feature_extraction}e) and thoracic aorta calcification mask (red regions in \cref{fig:discrete_feature_extraction}d and \cref{fig:discrete_feature_extraction}e) are directly tallied, and then multiplied by the voxel volume ($x$ resolution $\times$ $y$ resolution $\times$ $z$ resolution) to obtain CACV and ACV (in $mm^3$).

(iii) Based on the segmentation results of the thoracic aorta, three quantitative biomarkers characterizing the morphology of the thoracic aorta are computed: ATI, AMD, and AMDSTD. ATI represents the curvature of the thoracic aorta~\cite{girardi2021aortic}, while AMD and AMDSTD describe the overall maximum diameter of the aorta and its variation along the entire length. To calculate ATI, we first extract the centerline of the thoracic aorta using the segmentation outcome, and then determine the locations of the root point and end point (referred to as Point $R$ and Point $E$ in \cref{fig:discrete_feature_extraction}f). We then calculate ATI as the ratio of the length of the centerline (in $mm$) to the straight linear distance between the point $R$ and the point $E$ (also in $mm$). Subsequently, we generate cross-sectional views at 1 $mm$ intervals along the centerline, with \cref{fig:discrete_feature_extraction}i and \cref{fig:discrete_feature_extraction}j exemplifying the cross-sectional views generated at positions $n$ and $m$ in \cref{fig:discrete_feature_extraction}(f). For each cross-sectional view, we measure the maximum diameter of the thoracic aorta, denoted as $D=[d_1,...,d_n,...,d_m,...,d_k]$ (where $K$ = centerline length/interval (1 $mm$) + 1, it represents the total number of cross-sectional views). Finally, we compute AMD as the maximum value in $D$ (AMD = $max(D)$) and ANDSTD as the standard deviation of the diameters in $D$ (AMDSTD = $std(D)$).

(iv) Based on the results of the heart segmentation (cardiac chambers and pericardium), four quantitative biomarkers are computed to characterize the morphology and structure of the heart: CHR, CLD, CSD, and CTR~\cite{troxler2018role}. CHR is defined as the ratio of the volume of the cardiac chambers (the green regions in \cref{fig:discrete_feature_extraction}f) to the total volume of the entire heart (the combined red and green areas in \cref{fig:discrete_feature_extraction}f). Using the segmentation results of the heart, we select the maximal four-chambers axial view of the heart and perform an elliptical fitting on this segmented image to derive the cardiac long diameter (CLD) and short diameter (CSD). In \cref{fig:discrete_feature_extraction}f, half of line $AB$ represents the CSD while half of line $CD$ represents the CLD. Based on the calculation method described in the previous study~\cite{girardi2021aortic}, we determine the width of the heart, represented by line $FG$ in \cref{fig:discrete_feature_extraction}g. Similarly, utilizing the segmentation results of the lungs (as seen in \cref{fig:discrete_feature_extraction}l), we obtain the width of the entire lung, denoted by line $HI$ in \cref{fig:discrete_feature_extraction}g. Finally, CTR is calculated as the ratio of line $FG$ to line $HI$.

(v) The segmentation results of the lungs are used to calculate four quantitative biomarkers related to lung texture: LLR, RLR, LHR, and RHR. First, We use the segmentation outcomes for the left and right lungs separately to identify the regions of the lungs in the CT scan that correspond to pulmonary areas. We then further demarcate the regions with high attenuation intensity (as shown in \cref{fig:discrete_feature_extraction}m) using a threshold above -200HU. Consequently, LHR represents the ratio of the number of high attenuation intensity voxels to the total number of voxels in the left lung, and RHR is defined as the ratio of the number of high attenuation intensity voxels to the total number of voxels in the right lung. In a similar fashion, we delineate the regions with low attenuation intensity (as shown in \cref{fig:discrete_feature_extraction}n) using a threshold below -950HU, and then calculate the LLR and RLR respectively.

\begin{figure*}[t]
\renewcommand{\figurename}{Fig.}
\centering
\includegraphics[width=\textwidth]{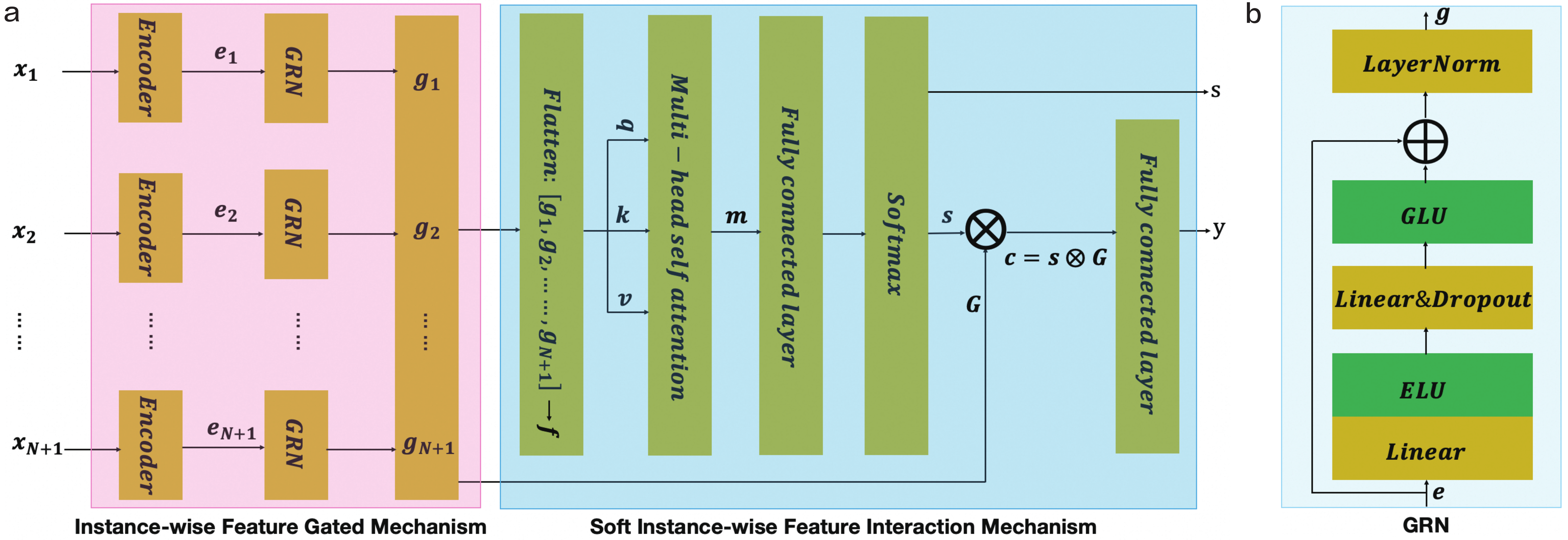}
\caption{\textbf{\textbar~The architecture of the features joint representation module and GRN block in DeepCVD.} \textbf{a.} Feature joint representation module. After passing through the instance-wise feature gated mechanism and soft instance-wise feature interaction mechanism, the input $X$ yield a CVD risk prediction result $y$ and contribution score $s$. \textbf{b.} Gated residual network (GRN) block.}
\label{fig:arch_net}
\end{figure*}

\noindent{\textbf{Continuous and discrete features joint representation}}

\noindent Recent studies have shown that methods that integrate features from various sources into one representation can offer complementary information to each other and lead to better performance~\cite{kiela2014learning}. However, not all features contribute to the prediction of the target, and certain artificially designed image quantitative biomarkers may actually hinder performance. Furthermore, the relevance and specific contributions of the features to the output target are typically unknown. Therefore, we have developed a features joint representation module (\cref{fig:arch_net}a) which can help identify which features are most significant for the prediction problem, reinforcing the most relevant features for CVD risk prediction, and suppressing any unnecessary features that could have a negative impact on performance. This module comprises an instance-wise feature-gated mechanism and a soft instance-wise feature interaction mechanism. 

\subsection{Instance-wise feature gated mechanism.} The high-dimensional continuous features are derived from a pre-trained deep model, while each discrete biomarker is a scalar feature quantified based on the prior knowledge of physicians. Therefore, the initial step is to ensure that these distinct types of features are properly aligned within the feature space. The extracted continuous and discrete features are denoted as $X=[x_1; x_2; ...; x_{N+1}]$. To maintain the relative independence of each type of feature and understand its contribution, we do not directly encode all features $X$. Instead, an individual encoder $F_i$ is applied to each $x_i$ as the input embedding $e_i$. The encoder $F_i$ consists of two fully connected layers:
\begin{equation}
\label{embedding_eq}
\begin{aligned}
e_i = F_i(W_i, x_i)
\end{aligned}
\end{equation}

where $e_i$ is a vector feature embedding of $x_i$ ($e_i \in \mathbb{R}^{1 \times L}$, in here $L=32$). $F_i$ is the $i^{th}$ encoding operation, and $W_i$ is the trainable weights of the encoder. Then we obtain the overall encoded features embedding matrix $E$ ($E \in \mathbb{R}^{{(N+1)} \times L}$).

To improve the expressive power of the model and better capture the relationship between features and the output target, we apply non-linear processing to each instance-wise feature embedding $e_i$. However, determining the extent of required non-linear processing remains a complex task. Some studies~\cite{pedro2000unified,hawkins2004problem} suggest that simpler models may benefit in datasets with noise. Given that our discrete quantitative biomarkers are automatically derived from the four pre-trained body part segmentation models without any additional adjustment, it implies that the input contains unavoidable noise. In consideration of this, we employ the GRN proposed in~\cite{lim2021temporal}, which is a notably flexible and simple architecture (\cref{fig:arch_net}b), as the non-linear operation.

\begin{equation}
\label{gated_residual_network_eq}
\begin{aligned}
GRN(e) = LayerNorm(GLU(\eta) + e)
\end{aligned}
\end{equation}
\begin{equation}
\label{elu_eq}
\begin{aligned}
\eta = Dropout(FC(ELU(FC(e))), p)
\end{aligned}
\end{equation}

Where $FC$ is the fully-connected Layer, $ELU$ is the Exponential Linear Unit activation function~\cite{clevert2015fast}, $GLU$ is the Gated Linear Units~\cite{dauphin2017language} and $LayerNorm$ is the standard layer normalization~\cite{ba2016layer}. During training, dropout is applied, and the dropout rate $p$ is set as 0.5. For each instance-wise feature embedding $e_i$, a non-linear operation is employed by its own GRN:
\begin{equation}
\label{gated_eq}
\begin{aligned}
g_i = GRN_i(e_i)
\end{aligned}
\end{equation}

\subsection{Soft instance-wise feature interaction mechanism.} Once the continuous and discrete features are aligned into the same dimensional space, we proceed to model the interactions between features and their contributions to the prediction target. The critical issue is how to fuse features from various sources to offer complementary information to each other. Given the recent significant performance of multi-head self-attention networks~\cite{vaswani2017attention} in modeling complex relationships, we utilize a multi-head attention mechanism to accomplish the interaction and fusion of features, employing different heads for different representation subspaces. First, we flatten the instance-wise feature embeddings $G=[g_1; g_2; ...; g_{N+1}]$ ($G \in \mathbb{R}^{{(N+1)} \times L}$) into a long-term relationship vector $f=[g_1^\intercal,g_2^\intercal,...,g_{N+1}^\intercal]^\intercal$, then we feed it to the multi-head attention module for feature interaction and combine outputs concatenated from all heads (in the study, head number set as 2):
\begin{equation}
\label{multi_head_attention}
\begin{aligned}
m = MultiHeadAttention(f)
\end{aligned}
\end{equation}

Finally, the instance-wise feature weights are generated by feeding $m$ through a fully connected layer, followed by a softmax layer:

\begin{equation}
\label{soft_selection_1}
\begin{aligned}
s = Softmax(FC(m))
\end{aligned}
\end{equation}

After we get the instance-wise feature contribution scores $s$ ($s \in \mathbb{R}^{1 \times {(N+1)}}$), then through the contribution scores $s$ and the processed instance-wise feature embeddings $G$, we can obtain the final classification representation $c$ ($c \in \mathbb{R}^{1 \times L}$):
\begin{equation}
\label{feature_weights}
\begin{aligned}
c = s \otimes G
\end{aligned}
\end{equation}
where $\otimes$ represents matrix inner-product operation.

\section*{Ablation studies}

We provide ablation analysis on the LDCT-NLST testing cohort to further investigate the effectiveness of the proposed joint representation approach.

\begin{figure*}[htbp]
\renewcommand{\figurename}{Fig.}
\centering
\includegraphics[width=\textwidth]{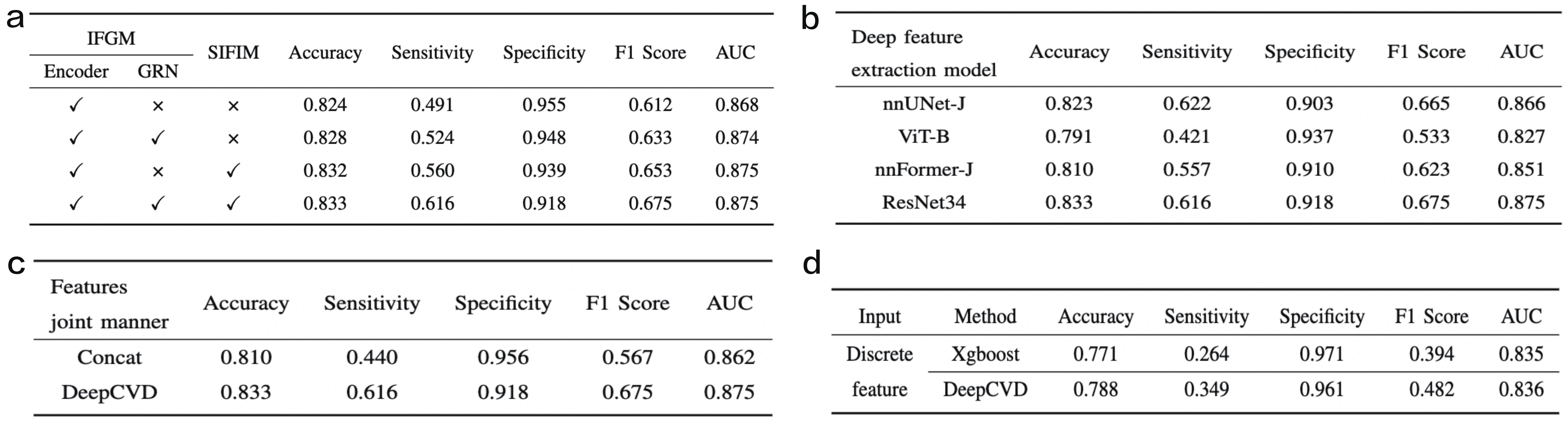}
\caption{\textbf{\textbar~An overview of the quantitative evaluation results from the ablation experiments conducted on the LDCT-NLST testing cohort.} \textbf{a.} Quantitative evaluation of the contribution of each component in DeepCVD. \textbf{b.} Depicts the impact of different deep feature extraction models at the first stage on CVD risk prediction performance; in DeepCVD, ResNet34 was employed as the deep feature extractor. \textbf{c.} Quantitative evaluation of DeepCVD with different features joint representation manners in the second stage. \textbf{d.} Quantitative evaluation of the effectiveness of features joint representation module in DeepCVD with only used discrete biomarkers as inputs.}
\label{fig:ablation_study}
\end{figure*}

\subsection{The effectiveness of different components.} To validate the effectiveness of various components in our proposed feature joint representation approach, we conducted an ablation study and presented the results on the LDCT-NLST dataset in \cref{fig:ablation_study}a. Several key observations can be drawn from these findings. Firstly, the incorporation of the GRN module significantly improves AUC, F1 Score, and Accuracy compared to using the Encoder alone when only the instance-wise feature-gated mechanism is employed. Secondly, even without using the GRN in the instance-wise feature-gated mechanism, the soft instance-wise feature interaction mechanism effectively boosts the model's performance. Thirdly, achieving considerable best results is possible when both the instance-wise feature gated mechanism and the soft instance-wise feature interaction mechanism are used simultaneously. This is primarily attributed to the efficient modeling of discrete features and deep continuous features, leveraging their respective strengths to promote the learning of the feature joint representation with the assistance of both the GRN and the soft instance-wise feature interaction mechanism. Additionally, our approach empirically confirms an intuitive phenomenon: as prior knowledge is more intricately processed within the model, its ability to enhance model specificity gradually decreases, helping the model to strike a balance between specificity and sensitivity.

\subsection{The impact of different deep feature extraction models.} We studied the influence of different deep feature extraction models on the performance of CVD risk prediction and presented the results in \cref{fig:ablation_study}b. It is observed that different deep feature extraction models have a certain impact on performance. Our approach demonstrates the best overall performance when employing the ResNet34 as the deep feature extractor, especially regarding AUC, Accuracy, and F1 Score. Furthermore, we also find that our approach performs better on most performance metrics when using light CNN architecture in the first stage than when using transformer architecture. For example, when using ResNet34 as opposed to ViT-B, the AUC increased from 0.827 to 0.875, an improvement of 5.8\%, Accuracy increased by 5.6\%, Sensitivity by 45.6\%, while Specificity only decreased by 1.6\%. Additionally, coupled with the data from \cref{tab:table_nlst}, it can be seen that regardless of which deep feature extraction model is used when combined with the discrete quantitative biomarkers for our proposed joint representational learning approach, their performance is enhanced to varying degrees. Specifically, ViT-B's AUC increased from 0.676 to 0.827, an improvement of 22.3\%, nnFormer's AUC from 0.837 to 0.851, and ResNet34's AUC from 0.844 to 0.875.

\subsection{The effectiveness of features joint representation module.} We validated the effectiveness of the feature joint representation module employed in the second stage from two perspectives: (i) Under the precondition of using the continuous and discrete quantitative biomarkers obtained from the first stage as inputs, our method demonstrates significant performance gains in Accuracy, Sensitivity, F1 Score, and AUC when compared with the approach that directly concatenates the features. Notably, Sensitivity improved by 40.0\%, and F1 Score by 19.0\%. Detailed results are shown in \cref{fig:ablation_study}c. (ii) When only the discrete quantitative biomarkers derived from the first stage are used as inputs, our joint representation approach substantially improves performance metrics including Accuracy, Sensitivity, F1 Score, and AUC over Xgboost, which also utilizes discrete quantitative biomarkers. The comprehensive results are presented in \cref{fig:ablation_study}d.

\section*{Comparison methods}

To comprehensively evaluate the effectiveness of our proposed method, we selected two different classes of methods for comparative analysis based on the various representations of features: (i) Machine learning methods utilizing discrete features, for which we chose the most commonly used Xgboost~\cite{chen2016Xgboost}. (ii) Deep learning methods using continuous features, including classification networks such as the CNN architecture ResNet34~\cite{he2016deep}, the CNN-Attention architecture Tri2D-Net~\cite{chao2021deep}, and the Transformer architecture ViT-B~\cite{dosovitskiy2020image}. Tri2D-Net is currently the leading method for CVD risk prediction on the LDCT-NLST dataset. The other category is based on multi-task learning approaches, with prior research~\cite{caruana1997multitask} demonstrating that models discover more general feature representations in solving multiple tasks, thereby improving generalization to unseen data. Additionally, we are aware that the nnUNet framework~\cite{isensee2021nnu} possesses strong data preprocessing and augmentation capabilities. Therefore, we integrated classification heads with fully connected layers and average pooling operations into segmentation networks such as nnUNet and nnFormer~\cite{zhou2023nnformer}. These heads receive input from the deepest feature layer of the encoder. Outputs provided by the multi-task methods include predicted cardiac segmentation masks and CVD risk classification probabilities. Except for Tri2D-Net, which directly uses an open-source model\footnote{https://github.com/DIAL-RPI/CVD-Risk-Estimator} trained on the LDCT-NLST training set, all other comparison methods are trained from scratch based on the LDCT-NLST training set.

\noindent{\textbf{Statistical analysis}}

\noindent The sensitivity and specificity of DeepCVD for CVD risk prediction were evaluated by calculating the 95\% confidence intervals using the Clopper-Pearson method based on 1,000 bootstrap replications of the data. In our setting, CVD risk prediction was a binary classification task, and p-values for accuracy comparisons were calculated through McNemar's test.

\section*{Reporting summary}
Further information on research design is available in the Nature Research Reporting Summary linked to this article.

\section*{Data availability}
The imaging data of this study are from a public chest low-dose CT dataset (LDCT-NLST) and a private external chest standard-dose CT dataset (NERC-MBD). The LDCT-NLST dataset is publicly available at \url{https://biometry.nci.
nih.gov/cdas/learn/nlst/images/}. The NERC-MBD dataset is used under a research agreement for the current study and is not publicly available. Source data is provided with this paper.

\section*{Code availability}
The code used for DeepCVD implementation depends on internal tooling and infrastructure, is under patent protection (application number: CN117274185B), and thus cannot be publicly released. All experiments and implementation details are described sufficiently in the Methods section for replication with non-proprietary libraries. The foundational architecture for the four specialized body part segmentation models of our work is available in an open source repository: \url{https://github.com/yhygao/CBIM-Medical-Image-Segmentation}. The ResNet34 continuous deep feature extraction model used in this study is implemented from: \url{https://github.com/pytorch}.

\section*{References}

\bibliographystyle{naturemag}
\bibliography{ref}

\begin{thebibliography}{10}
\expandafter\ifx\csname url\endcsname\relax
  \def\url#1{\texttt{#1}}\fi
\expandafter\ifx\csname urlprefix\endcsname\relax\def\urlprefix{URL }\fi
\providecommand{\bibinfo}[2]{#2}
\providecommand{\eprint}[2][]{\url{#2}}

\bibitem{wilkins2017european}
\bibinfo{author}{Wilkins, E.} \emph{et~al.}
\newblock \bibinfo{title}{European cardiovascular disease statistics 2017}  (\bibinfo{year}{2017}).

\bibitem{mozaffarian2016heart}
\bibinfo{author}{Mozaffarian, D.} \emph{et~al.}
\newblock \bibinfo{title}{Heart disease and stroke statistics—2016 update: a report from the american heart association}.
\newblock \emph{\bibinfo{journal}{circulation}} \textbf{\bibinfo{volume}{133}}, \bibinfo{pages}{e38--e360} (\bibinfo{year}{2016}).

\bibitem{d20191997}
\bibinfo{author}{D'Souza, M.~J.}, \bibinfo{author}{Li, R.~C.}, \bibinfo{author}{Gannon, M.~L.} \& \bibinfo{author}{Wentzien, D.~E.}
\newblock \bibinfo{title}{1997--2017 leading causes of death information due to diabetes, neoplasms, and diseases of the circulatory system, issues cautionary weight-related lesson to the us population at large}.
\newblock In \emph{\bibinfo{booktitle}{2019 International Conference on Engineering, Science, and Industrial Applications (ICESI)}}, \bibinfo{pages}{1--6} (\bibinfo{organization}{IEEE}, \bibinfo{year}{2019}).

\bibitem{ridker2008rosuvastatin}
\bibinfo{author}{Ridker, P.~M.} \emph{et~al.}
\newblock \bibinfo{title}{Rosuvastatin to prevent vascular events in men and women with elevated c-reactive protein}.
\newblock \emph{\bibinfo{journal}{New England journal of medicine}} \textbf{\bibinfo{volume}{359}}, \bibinfo{pages}{2195--2207} (\bibinfo{year}{2008}).

\bibitem{greenland20102010}
\bibinfo{author}{Greenland, P.} \emph{et~al.}
\newblock \bibinfo{title}{2010 accf/aha guideline for assessment of cardiovascular risk in asymptomatic adults: a report of the american college of cardiology foundation/american heart association task force on practice guidelines developed in collaboration with the american society of echocardiography, american society of nuclear cardiology, society of atherosclerosis imaging and prevention, society for cardiovascular angiography and interventions, society of cardiovascular computed tomography, and society for cardiovascular magnetic resonance}.
\newblock \emph{\bibinfo{journal}{Journal of the American College of Cardiology}} \textbf{\bibinfo{volume}{56}}, \bibinfo{pages}{e50--e103} (\bibinfo{year}{2010}).

\bibitem{wilson1998prediction}
\bibinfo{author}{Wilson, P.~W.} \emph{et~al.}
\newblock \bibinfo{title}{Prediction of coronary heart disease using risk factor categories}.
\newblock \emph{\bibinfo{journal}{Circulation}} \textbf{\bibinfo{volume}{97}}, \bibinfo{pages}{1837--1847} (\bibinfo{year}{1998}).

\bibitem{eichler2007prediction}
\bibinfo{author}{Eichler, K.}, \bibinfo{author}{Puhan, M.~A.}, \bibinfo{author}{Steurer, J.} \& \bibinfo{author}{Bachmann, L.~M.}
\newblock \bibinfo{title}{Prediction of first coronary events with the framingham score: a systematic review}.
\newblock \emph{\bibinfo{journal}{American heart journal}} \textbf{\bibinfo{volume}{153}}, \bibinfo{pages}{722--731} (\bibinfo{year}{2007}).

\bibitem{d2008general}
\bibinfo{author}{D’Agostino~Sr, R.~B.} \emph{et~al.}
\newblock \bibinfo{title}{General cardiovascular risk profile for use in primary care: the framingham heart study}.
\newblock \emph{\bibinfo{journal}{Circulation}} \textbf{\bibinfo{volume}{117}}, \bibinfo{pages}{743--753} (\bibinfo{year}{2008}).

\bibitem{piepoli2016guidelines}
\bibinfo{author}{Piepoli, M.~F.} \emph{et~al.}
\newblock \bibinfo{title}{Guidelines: Editor's choice: 2016 european guidelines on cardiovascular disease prevention in clinical practice: The sixth joint task force of the european society of cardiology and other societies on cardiovascular disease prevention in clinical practice (constituted by representatives of 10 societies and by invited experts) developed with the special contribution of the european association for cardiovascular prevention \& rehabilitation (eacpr)}.
\newblock \emph{\bibinfo{journal}{European heart journal}} \textbf{\bibinfo{volume}{37}}, \bibinfo{pages}{2315} (\bibinfo{year}{2016}).

\bibitem{yang2016predicting}
\bibinfo{author}{Yang, X.} \emph{et~al.}
\newblock \bibinfo{title}{Predicting the 10-year risks of atherosclerotic cardiovascular disease in chinese population: the china-par project (prediction for ascvd risk in china)}.
\newblock \emph{\bibinfo{journal}{Circulation}} \textbf{\bibinfo{volume}{134}}, \bibinfo{pages}{1430--1440} (\bibinfo{year}{2016}).

\bibitem{liu2018predicting}
\bibinfo{author}{Liu, F.} \emph{et~al.}
\newblock \bibinfo{title}{Predicting lifetime risk for developing atherosclerotic cardiovascular disease in chinese population: the china-par project}.
\newblock \emph{\bibinfo{journal}{Science bulletin}} \textbf{\bibinfo{volume}{63}}, \bibinfo{pages}{779--787} (\bibinfo{year}{2018}).

\bibitem{pickhardt2020automated}
\bibinfo{author}{Pickhardt, P.~J.} \emph{et~al.}
\newblock \bibinfo{title}{Automated ct biomarkers for opportunistic prediction of future cardiovascular events and mortality in an asymptomatic screening population: a retrospective cohort study}.
\newblock \emph{\bibinfo{journal}{The Lancet Digital Health}} \textbf{\bibinfo{volume}{2}}, \bibinfo{pages}{e192--e200} (\bibinfo{year}{2020}).

\bibitem{xu2023ai}
\bibinfo{author}{Xu, K.} \emph{et~al.}
\newblock \bibinfo{title}{Ai body composition in lung cancer screening: added value beyond lung cancer detection}.
\newblock \emph{\bibinfo{journal}{Radiology}} \textbf{\bibinfo{volume}{308}}, \bibinfo{pages}{e222937} (\bibinfo{year}{2023}).

\bibitem{eng2021automated}
\bibinfo{author}{Eng, D.} \emph{et~al.}
\newblock \bibinfo{title}{Automated coronary calcium scoring using deep learning with multicenter external validation}.
\newblock \emph{\bibinfo{journal}{NPJ digital medicine}} \textbf{\bibinfo{volume}{4}}, \bibinfo{pages}{88} (\bibinfo{year}{2021}).

\bibitem{zeleznik2021deep}
\bibinfo{author}{Zeleznik, R.} \emph{et~al.}
\newblock \bibinfo{title}{Deep convolutional neural networks to predict cardiovascular risk from computed tomography}.
\newblock \emph{\bibinfo{journal}{Nature communications}} \textbf{\bibinfo{volume}{12}}, \bibinfo{pages}{715} (\bibinfo{year}{2021}).

\bibitem{siontis2012comparisons}
\bibinfo{author}{Siontis, G.~C.}, \bibinfo{author}{Tzoulaki, I.}, \bibinfo{author}{Siontis, K.~C.} \& \bibinfo{author}{Ioannidis, J.~P.}
\newblock \bibinfo{title}{Comparisons of established risk prediction models for cardiovascular disease: systematic review}.
\newblock \emph{\bibinfo{journal}{Bmj}} \textbf{\bibinfo{volume}{344}} (\bibinfo{year}{2012}).

\bibitem{van2019direct}
\bibinfo{author}{van Velzen, S.~G.} \emph{et~al.}
\newblock \bibinfo{title}{Direct prediction of cardiovascular mortality from low-dose chest ct using deep learning}.
\newblock In \emph{\bibinfo{booktitle}{Medical Imaging 2019: Image Processing}}, vol. \bibinfo{volume}{10949}, \bibinfo{pages}{240--245} (\bibinfo{organization}{SPIE}, \bibinfo{year}{2019}).

\bibitem{chao2021deep}
\bibinfo{author}{Chao, H.} \emph{et~al.}
\newblock \bibinfo{title}{Deep learning predicts cardiovascular disease risks from lung cancer screening low dose computed tomography}.
\newblock \emph{\bibinfo{journal}{Nature Communications}} \textbf{\bibinfo{volume}{12}}, \bibinfo{pages}{2963} (\bibinfo{year}{2021}).

\bibitem{lim2021temporal}
\bibinfo{author}{Lim, B.}, \bibinfo{author}{Ar{\i}k, S.~{\"O}.}, \bibinfo{author}{Loeff, N.} \& \bibinfo{author}{Pfister, T.}
\newblock \bibinfo{title}{Temporal fusion transformers for interpretable multi-horizon time series forecasting}.
\newblock \emph{\bibinfo{journal}{International Journal of Forecasting}} \textbf{\bibinfo{volume}{37}}, \bibinfo{pages}{1748--1764} (\bibinfo{year}{2021}).

\bibitem{vaswani2017attention}
\bibinfo{author}{Vaswani, A.} \emph{et~al.}
\newblock \bibinfo{title}{Attention is all you need}.
\newblock \emph{\bibinfo{journal}{Advances in neural information processing systems}} \textbf{\bibinfo{volume}{30}} (\bibinfo{year}{2017}).

\bibitem{chen2016Xgboost}
\bibinfo{author}{Chen, T.} \& \bibinfo{author}{Guestrin, C.}
\newblock \bibinfo{title}{Xgboost: A scalable tree boosting system}.
\newblock In \emph{\bibinfo{booktitle}{Proceedings of the 22nd acm sigkdd international conference on knowledge discovery and data mining}}, \bibinfo{pages}{785--794} (\bibinfo{year}{2016}).

\bibitem{he2016deep}
\bibinfo{author}{He, K.}, \bibinfo{author}{Zhang, X.}, \bibinfo{author}{Ren, S.} \& \bibinfo{author}{Sun, J.}
\newblock \bibinfo{title}{Deep residual learning for image recognition}.
\newblock In \emph{\bibinfo{booktitle}{Proceedings of the IEEE conference on computer vision and pattern recognition}}, \bibinfo{pages}{770--778} (\bibinfo{year}{2016}).

\bibitem{van2008visualizing}
\bibinfo{author}{Van~der Maaten, L.} \& \bibinfo{author}{Hinton, G.}
\newblock \bibinfo{title}{Visualizing data using t-sne.}
\newblock \emph{\bibinfo{journal}{Journal of machine learning research}} \textbf{\bibinfo{volume}{9}} (\bibinfo{year}{2008}).

\bibitem{kiela2014learning}
\bibinfo{author}{Kiela, D.} \& \bibinfo{author}{Bottou, L.}
\newblock \bibinfo{title}{Learning image embeddings using convolutional neural networks for improved multi-modal semantics}.
\newblock In \emph{\bibinfo{booktitle}{Proceedings of the 2014 Conference on empirical methods in natural language processing (EMNLP)}}, \bibinfo{pages}{36--45} (\bibinfo{year}{2014}).

\bibitem{selvaraju2017grad}
\bibinfo{author}{Selvaraju, R.~R.} \emph{et~al.}
\newblock \bibinfo{title}{Grad-cam: Visual explanations from deep networks via gradient-based localization}.
\newblock In \emph{\bibinfo{booktitle}{Proceedings of the IEEE international conference on computer vision}}, \bibinfo{pages}{618--626} (\bibinfo{year}{2017}).

\bibitem{huang2020fusion}
\bibinfo{author}{Huang, S.-C.}, \bibinfo{author}{Pareek, A.}, \bibinfo{author}{Seyyedi, S.}, \bibinfo{author}{Banerjee, I.} \& \bibinfo{author}{Lungren, M.~P.}
\newblock \bibinfo{title}{Fusion of medical imaging and electronic health records using deep learning: a systematic review and implementation guidelines}.
\newblock \emph{\bibinfo{journal}{NPJ digital medicine}} \textbf{\bibinfo{volume}{3}}, \bibinfo{pages}{136} (\bibinfo{year}{2020}).

\bibitem{iacobellis2022epicardial}
\bibinfo{author}{Iacobellis, G.}
\newblock \bibinfo{title}{Epicardial adipose tissue in contemporary cardiology}.
\newblock \emph{\bibinfo{journal}{Nature reviews cardiology}} \textbf{\bibinfo{volume}{19}}, \bibinfo{pages}{593--606} (\bibinfo{year}{2022}).

\bibitem{gao2022data}
\bibinfo{author}{Gao, Y.} \emph{et~al.}
\newblock \bibinfo{title}{A data-scalable transformer for medical image segmentation: architecture, model efficiency, and benchmark}.
\newblock \emph{\bibinfo{journal}{arXiv preprint arXiv:2203.00131}}  (\bibinfo{year}{2022}).

\bibitem{girardi2021aortic}
\bibinfo{author}{Girardi, L.~N.}, \bibinfo{author}{Lau, C.} \& \bibinfo{author}{Gambardella, I.}
\newblock \bibinfo{title}{Aortic dimensions as predictors of adverse events}.
\newblock \emph{\bibinfo{journal}{The Journal of Thoracic and Cardiovascular Surgery}} \textbf{\bibinfo{volume}{161}}, \bibinfo{pages}{1193--1197} (\bibinfo{year}{2021}).

\bibitem{hemingway1998cardiothoracic}
\bibinfo{author}{Hemingway, H.}, \bibinfo{author}{Shipley, M.}, \bibinfo{author}{Christie, D.} \& \bibinfo{author}{Marmot, M.}
\newblock \bibinfo{title}{Cardiothoracic ratio and relative heart volume as predictors of coronary heart disease mortality: The whitehall study 25 year follow-up}.
\newblock \emph{\bibinfo{journal}{European heart journal}} \textbf{\bibinfo{volume}{19}}, \bibinfo{pages}{859--869} (\bibinfo{year}{1998}).

\bibitem{dey2012epicardial}
\bibinfo{author}{Dey, D.}, \bibinfo{author}{Nakazato, R.}, \bibinfo{author}{Li, D.} \& \bibinfo{author}{Berman, D.~S.}
\newblock \bibinfo{title}{Epicardial and thoracic fat-noninvasive measurement and clinical implications}.
\newblock \emph{\bibinfo{journal}{Cardiovascular diagnosis and therapy}} \textbf{\bibinfo{volume}{2}}, \bibinfo{pages}{85} (\bibinfo{year}{2012}).

\bibitem{agatston1990quantification}
\bibinfo{author}{Agatston, A.~S.} \emph{et~al.}
\newblock \bibinfo{title}{Quantification of coronary artery calcium using ultrafast computed tomography}.
\newblock \emph{\bibinfo{journal}{Journal of the American college of cardiology}} \textbf{\bibinfo{volume}{15}}, \bibinfo{pages}{827--832} (\bibinfo{year}{1990}).

\bibitem{troxler2018role}
\bibinfo{author}{Troxler, R.} \emph{et~al.}
\newblock \bibinfo{title}{The role of angiography in the congruence of cardiovascular measurements between autopsy and postmortem imaging}.
\newblock \emph{\bibinfo{journal}{International journal of legal medicine}} \textbf{\bibinfo{volume}{132}}, \bibinfo{pages}{249--262} (\bibinfo{year}{2018}).

\bibitem{pedro2000unified}
\bibinfo{author}{Pedro, D.}
\newblock \bibinfo{title}{A unified bias-variance decomposition and its applications}.
\newblock In \emph{\bibinfo{booktitle}{17th International Conference on Machine Learning}}, \bibinfo{pages}{231--238} (\bibinfo{year}{2000}).

\bibitem{hawkins2004problem}
\bibinfo{author}{Hawkins, D.~M.}
\newblock \bibinfo{title}{The problem of overfitting}.
\newblock \emph{\bibinfo{journal}{Journal of chemical information and computer sciences}} \textbf{\bibinfo{volume}{44}}, \bibinfo{pages}{1--12} (\bibinfo{year}{2004}).

\bibitem{clevert2015fast}
\bibinfo{author}{Clevert, D.-A.}, \bibinfo{author}{Unterthiner, T.} \& \bibinfo{author}{Hochreiter, S.}
\newblock \bibinfo{title}{Fast and accurate deep network learning by exponential linear units (elus)}.
\newblock \emph{\bibinfo{journal}{arXiv preprint arXiv:1511.07289}}  (\bibinfo{year}{2015}).

\bibitem{dauphin2017language}
\bibinfo{author}{Dauphin, Y.~N.}, \bibinfo{author}{Fan, A.}, \bibinfo{author}{Auli, M.} \& \bibinfo{author}{Grangier, D.}
\newblock \bibinfo{title}{Language modeling with gated convolutional networks}.
\newblock In \emph{\bibinfo{booktitle}{International conference on machine learning}}, \bibinfo{pages}{933--941} (\bibinfo{organization}{PMLR}, \bibinfo{year}{2017}).

\bibitem{ba2016layer}
\bibinfo{author}{Ba, J.~L.}, \bibinfo{author}{Kiros, J.~R.} \& \bibinfo{author}{Hinton, G.~E.}
\newblock \bibinfo{title}{Layer normalization}.
\newblock \emph{\bibinfo{journal}{arXiv preprint arXiv:1607.06450}}  (\bibinfo{year}{2016}).

\bibitem{dosovitskiy2020image}
\bibinfo{author}{Dosovitskiy, A.} \emph{et~al.}
\newblock \bibinfo{title}{An image is worth 16x16 words: Transformers for image recognition at scale}.
\newblock \emph{\bibinfo{journal}{arXiv preprint arXiv:2010.11929}}  (\bibinfo{year}{2020}).

\bibitem{caruana1997multitask}
\bibinfo{author}{Caruana, R.}
\newblock \bibinfo{title}{Multitask learning}.
\newblock \emph{\bibinfo{journal}{Machine learning}} \textbf{\bibinfo{volume}{28}}, \bibinfo{pages}{41--75} (\bibinfo{year}{1997}).

\bibitem{isensee2021nnu}
\bibinfo{author}{Isensee, F.}, \bibinfo{author}{Jaeger, P.~F.}, \bibinfo{author}{Kohl, S.~A.}, \bibinfo{author}{Petersen, J.} \& \bibinfo{author}{Maier-Hein, K.~H.}
\newblock \bibinfo{title}{nnu-net: a self-configuring method for deep learning-based biomedical image segmentation}.
\newblock \emph{\bibinfo{journal}{Nature methods}} \textbf{\bibinfo{volume}{18}}, \bibinfo{pages}{203--211} (\bibinfo{year}{2021}).

\bibitem{zhou2023nnformer}
\bibinfo{author}{Zhou, H.-Y.} \emph{et~al.}
\newblock \bibinfo{title}{nnformer: volumetric medical image segmentation via a 3d transformer}.
\newblock \emph{\bibinfo{journal}{IEEE Transactions on Image Processing}}  (\bibinfo{year}{2023}).

\end{thebibliography}

\section*{Acknowledgments} 

\section*{Author contributions}
For the three first co-authors, M.X. was responsible for data cleaning, deep learning model development, internal evaluation and drafted the manuscript, C.F. was responsible for collecting and preprocessing external data, as well as conducting evaluations of various methods on the external data, Y.Z. was responsible for the model training of different comparison methods, and they all participated in the experimental design and drafted the manuscript. W.G. helped deploy all test models to evaluate the external data. J.Q. and P.L. helped collect the external testing cohort. L.L. aided in the experimental design and edited the manuscript. H.C. assisted with the evaluation of the models. M.X. and K.H. were responsible for the conception and design of the experiments and oversaw overall direction and planning and drafted the manuscript.

\section*{Competing interests}
The authors declare that they have no competing interests.

\setcounter{figure}{0}

\begin{figure*}[htbp]
\renewcommand{\figurename}{Extended Data Fig.}
\centering
\includegraphics[width=\textwidth]{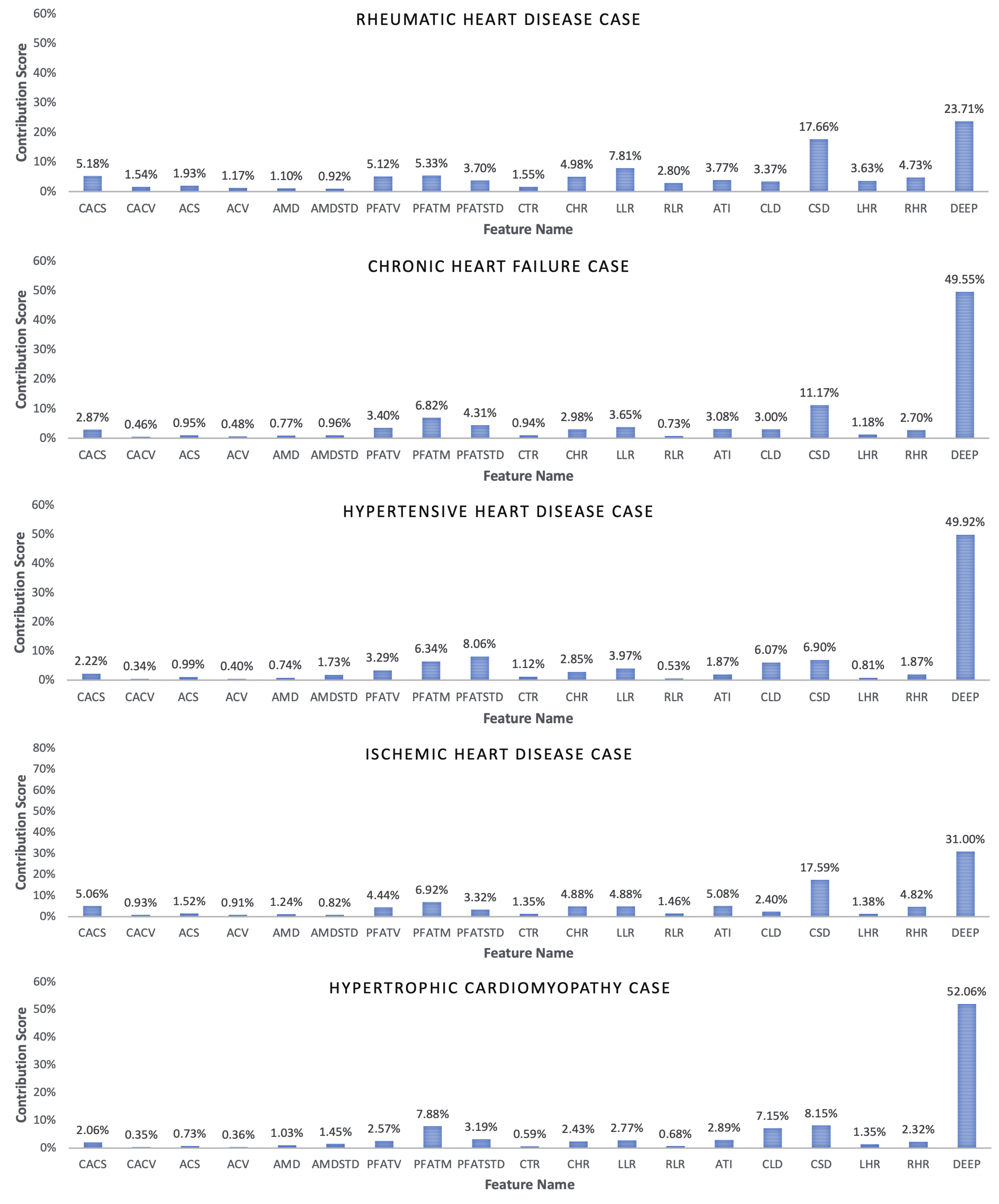}
\caption{\textbf{\textbar~Visualization of contribution score for different heart diseases.} From top to bottom, there are rheumatic heart disease, chronic heart failure, hypertensive heart disease, ischemic heart disease, and hypertrophic cardiomyopathy. They exhibit very similar contribution scores.}
\label{fig:contribution_score_s_1}
\end{figure*}

\begin{figure*}[htbp]
\renewcommand{\figurename}{Extended Data Fig.}
\centering
\includegraphics[width=\textwidth]{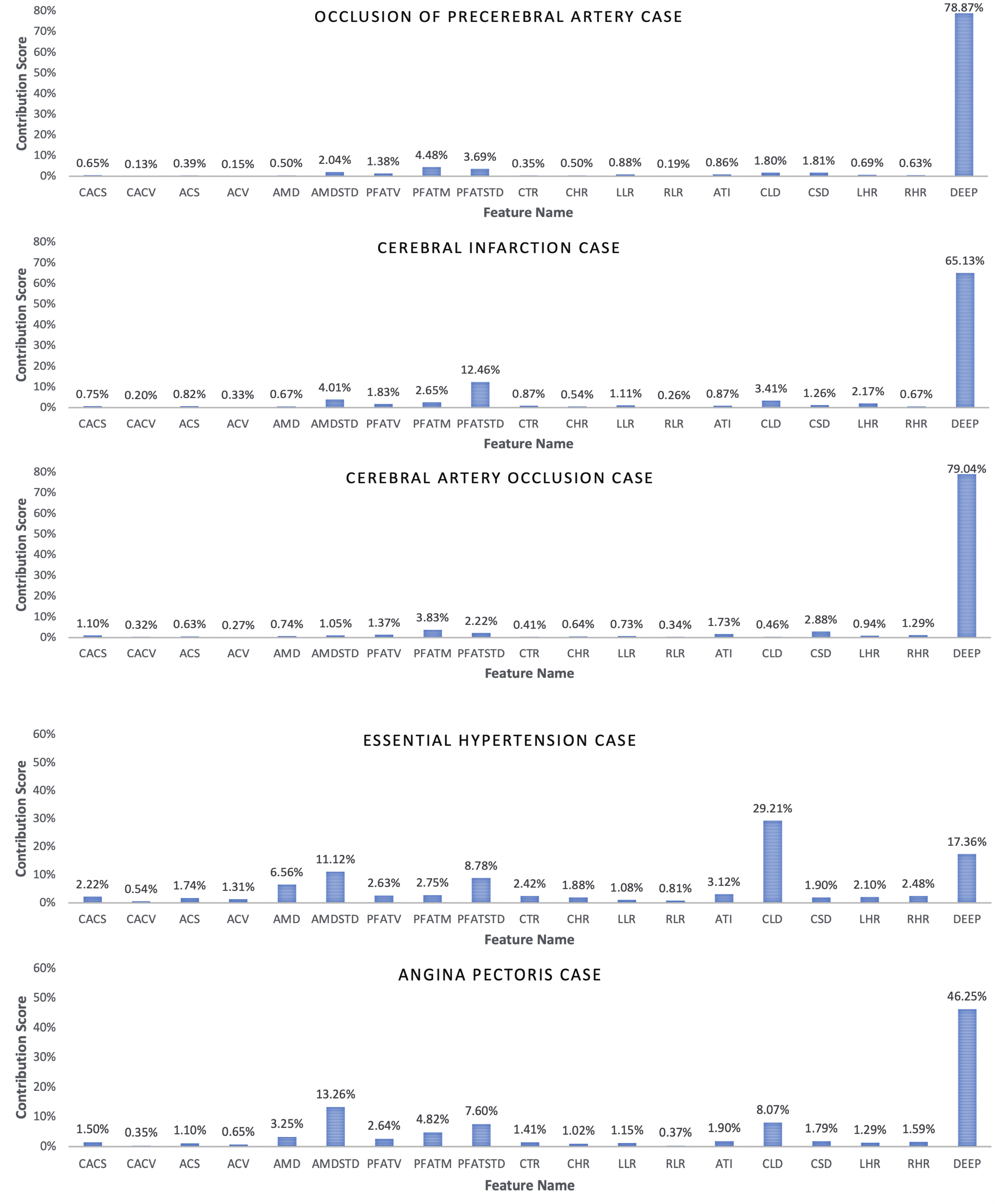}
\caption{\textbf{\textbar~More visualization of contribution score for different CVDs.} From top to bottom: occlusion of the precerebral artery, cerebral infarction, cerebral arterial occlusion, essential hypertension, and angina pectoris. It is seen that deep features play a dominant role in cerebrovascular-related diseases.}
\label{fig:contribution_score_s_2}
\end{figure*}

\begin{figure*}[htbp]
\renewcommand{\figurename}{Extended Data Fig.}
\centering
\includegraphics[width=\textwidth]{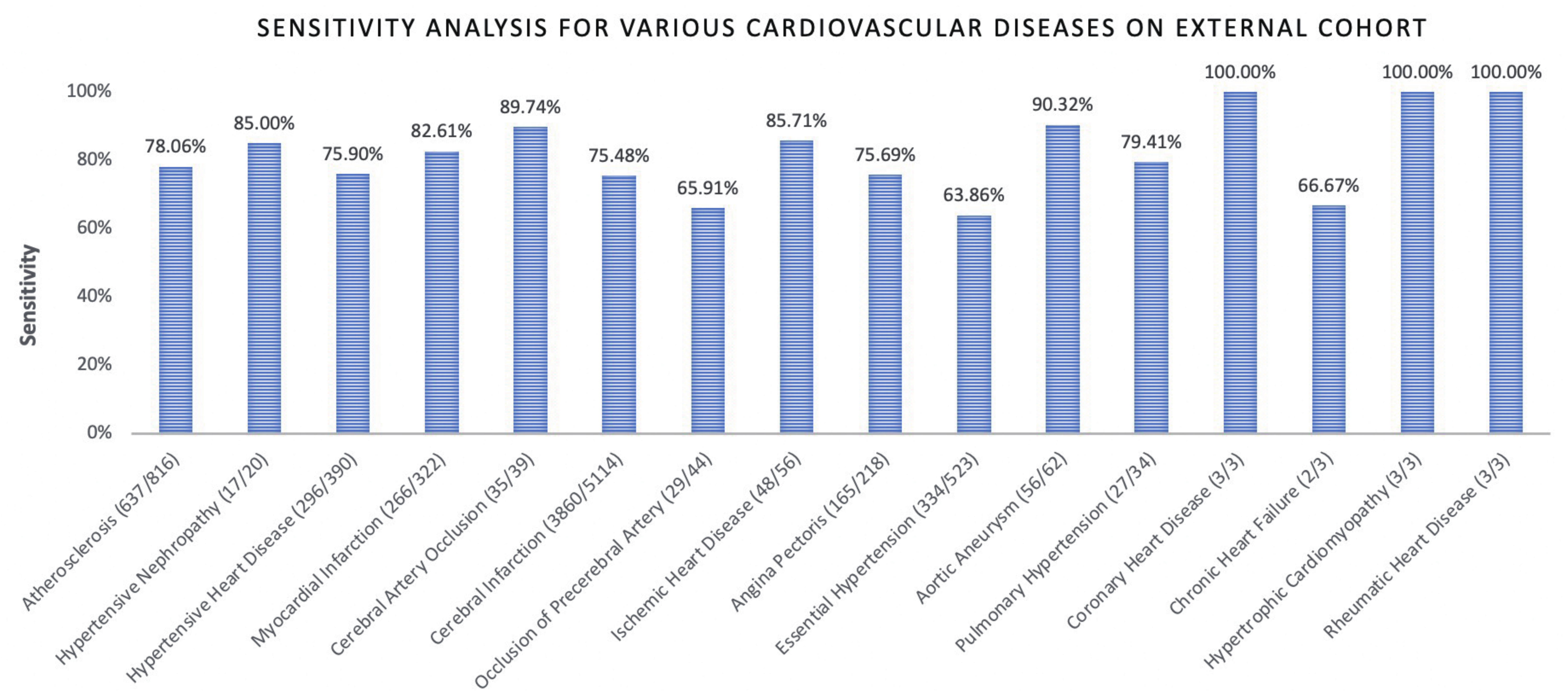}
\caption{\textbf{\textbar~Seneitivity of DeepCVD across 16 CVDs in the external testing cohort (NERC-MBD) of 7,650 CVD-Positive CT volumes.} Additionally, it is important to note that, for instance, in ``Atherosclerosis (637/816)'', ``Atherosclerosis'' represents the name of the disease, ``637'' denotes the number of CT volumes where DeepCVD predicted CVD-Positive, and ``816'' refers to the total number of CT volumes with Atherosclerosis in the cohort.}
\label{fig:cvd_diseases_subtypes}
\end{figure*}

\end{document}